\documentclass[%
 aip,
 amsmath,amssymb,
 reprint,%
]{revtex4-1}

\usepackage{graphicx}
\usepackage{dcolumn}
\usepackage{bm}
\usepackage{xcolor}
\usepackage[utf8]{inputenc}
\usepackage[T1]{fontenc}
\usepackage{mathptmx}

\begin{document}

\preprint{AIP/123-QED}

\title[Fast algorithm to identify cluster synchrony through fibration symmetries in large information-processing networks]{Fast algorithm to identify cluster synchrony through fibration symmetries in large information-processing networks}

\author{Higor S. Monteiro}
\affiliation{Departamento de F\'{i}sica, Universidade Federal do Cear\'{a}, Fortaleza, Cear\'{a}, Brazil 60451-970}
\author{Ian Leifer}%
\affiliation{Levich Institute and Physics Department, The City College of New York, New York, NY, USA 10031}%
\author{Saulo D. S. Reis}
\affiliation{Departamento de F\'{i}sica, Universidade Federal do Cear\'{a}, Fortaleza, Cear\'{a}, Brazil 60451-970}
\author{Jos\'{e} S. Andrade, Jr.}
\affiliation{Departamento de F\'{i}sica, Universidade Federal do Cear\'{a}, Fortaleza, Cear\'{a}, Brazil 60451-970}
\author{Hernan A. Makse}
\affiliation{Levich Institute and Physics Department, The City College of New York, New York, NY, USA 10031}%

\date{\today}

\begin{abstract}
    Recent studies revealed an important interplay between the detailed structure of fibration symmetric circuits and the functionality of biological and non-biological networks within which they have be identified. The presence of these circuits in complex networks are directed related to the phenomenon of cluster synchronization, which produces patterns of synchronized group of nodes.
    Here we present a fast, and memory efficient, algorithm to identify fibration symmetries over information-processing networks. This algorithm is specially suitable for large and sparse networks since it has runtime of complexity $\mathcal{O}(M\log N)$ and requires $\mathcal{O}(M+N)$ of memory resources, where $N$ and $M$ are the number of nodes and edges in the network, respectively. 
    We propose a modification on the so-called refinement paradigm to identify circuits symmetrical to information flow ({\it i.e.}, fibers) by finding the coarsest refinement partition over the network. 
    Finally, we show that the presented algorithm provides an optimal procedure for identifying fibers, overcoming the current approaches used in the literature.
\end{abstract}

\maketitle

\begin{quotation}
    Network fibers are circuits with fibration symmetries. These symmetries imply that nodes sharing a fiber receive identical information from the rest of the network, leading to the synchronization of their dynamics. Since these symmetries offer a conceptual approach to identify functional building blocks in networks, in contrast to network motifs \cite{motifs2002}, it is desirable to develop an efficient algorithm capable of extracting these circuits for large networks. Different methods have been proposed for the identification of fibers and, in special, an example of a balanced coloring algorithm has been used in the recent work of Morone \textit{et. al.} \cite{fibration2019} to show the ubiquitousness of these symmetries across several real networks. However, even though these methods are intuitive and relatively simple to implement, they can be highly inefficient regarding either their time or space complexity, limiting their applications to small networks. In this work, we show that refinement algorithms represent a natural approach to identify fibration symmetries, allowing to build an efficient method to find fibration building blocks in very large sparse networks. 
\end{quotation}

\section{\label{sec:level1}Introduction}

The progress of network science in the last two decades has shown an insightful framework for theoretical and practical studies regarding dynamical systems on networks~\cite{Strogatz2001, Bocaletti2006, DynamicsNet2016}. Dynamical processes constrained by network structures can display novel qualitative behaviors which are not directly observed in classical dynamical systems~\cite{Stewart2006}. This is specially true for information-processing systems, in which the phenomenon of synchronization~\cite{Arenas2008} is directly related to functionality.

Recently, Morone and collaborators~\cite{fibration2019} used the formalism of fibration symmetries on graphs~\cite{Boldi2002} to characterize sets of synchronized nodes on several complex networks. These sets, known as {\it fibers}, can provide vital information about the structure-function relation of networks to which they belong, since broken fibration symmetries allow fibers to behave as logical computational circuits~\cite{transistor2019}. From these novel structures, it is possible to define states of synchrony by considering only the topological features of the given network~\cite{pecora2014}. The concept of fibration symmetry provides a framework for the identification of these states. This approach comes from the recognition that, instead of restrictive conditions imposed by topological isomorphism, synchronization of information processing obeys a less constraining symmetry represented by the rules of groupoids~\cite{Stewart2006}. Specifically, information flow in directed networks is invariant under a graph fibration morphism~\cite{Boldi2006}, where a graph fibration is any transformation keeping invariant the set of input trees~\cite{fibration2019}.

For a proper definition of input tree, we first define the concept of input set of a node $v \in V$ of the network $G(V, E)$. A input set of a node $v$ is the set of pairs $( \text{type}(e_G), w)$ such that $s(e_G) = w$ and $t(e_G) = v$, where $s(e_G)$ and $t(e_G)$ are, respectively, the source and the target of the edge $e_G \in E$, and $w \in V$. If the network is a multiplex, meaning that there are more than one type of edge, the $\text{type}(e_G)$ is a necessary information to define correctly the input set of $v$. Otherwise, if the network contains only one type of edge, the input set of $v$ reduces to all its incoming neighbors $w$. Following this definition, we can define recursively an input tree by accessing input sets of input sets through a prescribed amount of steps. This recursive procedure is infinite if there are cycles between input sets.

\begin{figure}[t]
  \centering
  \includegraphics[scale=0.225]{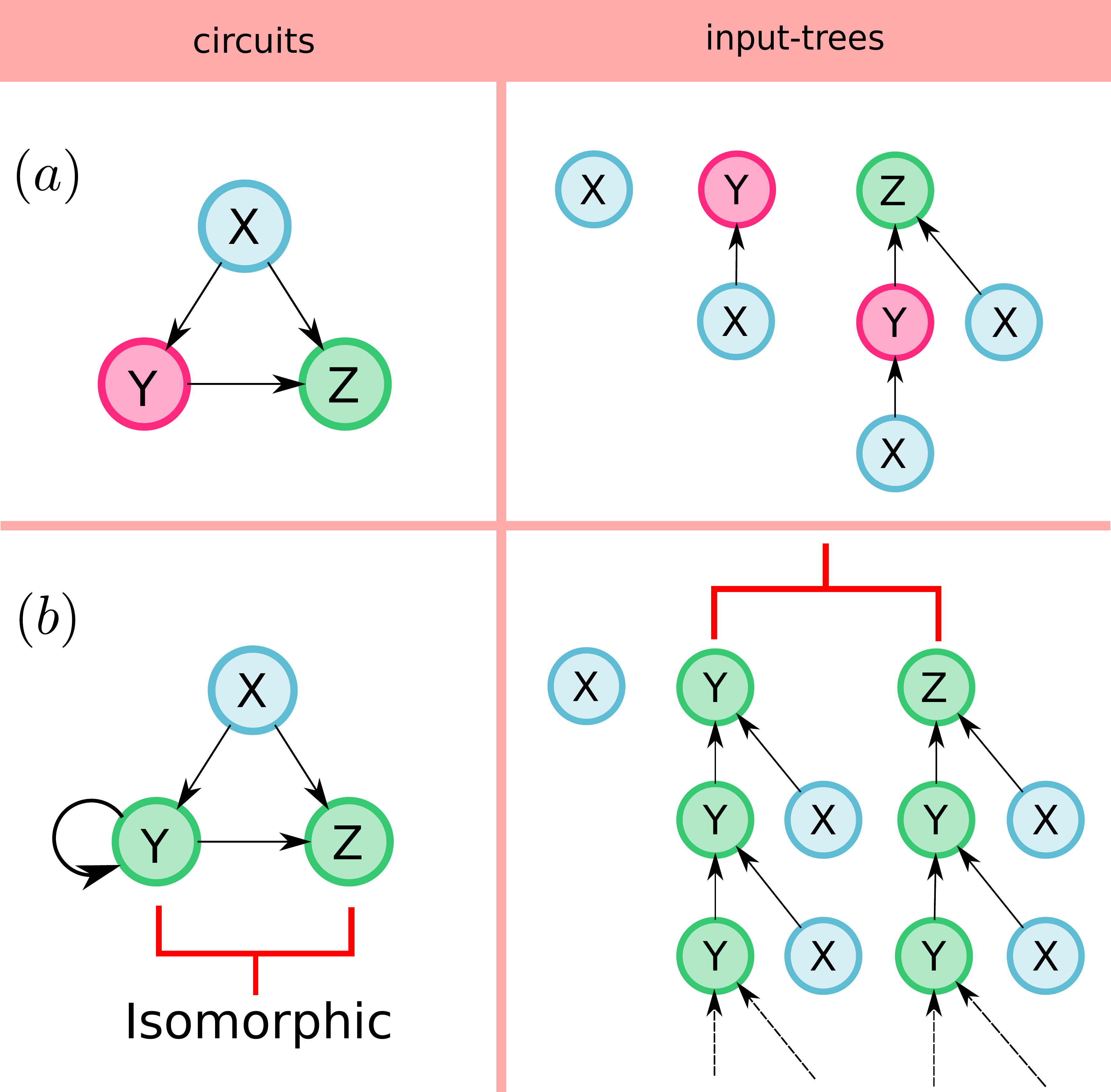}
  \caption{\textbf{Example of non-symmetric and symmetric circuits}. In (a) the input tree of each node differs from the others, showing non-equivalence on the information flow received by each node. As shown in (b), by adding a simple self-loop to node Y, its input tree becomes isomorphic to the input tree of Z. From this, their incoming information flows become equivalent, allowing both nodes to synchronize in their dynamics.}
  \label{fig:ex}
\end{figure}

Intuitively, to fully characterize the information flow paths in a directed network, we can define for each node $v$ an input tree containing all the information pathways that terminate on $v$. Thus, if two different nodes have isomorphic input trees, indicating that they receive equivalent information from the network, they synchronize their dynamical states, exhibiting a correlated dynamics in the network. If the network is composed by different groups of synchronized nodes, we say that there is a cluster synchronization of the dynamics on this network~\cite{pecora2014}. Moreover, if a group of nodes have isomorphic input trees, then a graph fibration morphism collapses these nodes from the network $G$ into one single node of a representative \textit{base} network $B$. In this way, a fibration reduces the original network to a base network where each node in $B$ represents a fiber module of $G$ in which all nodes inside this module have isomorphic input trees. Examples of directed networks formed by three nodes are shown in Fig.~\ref{fig:ex}. Here, we show that these different patterns of connections are responsible for the different cluster synchronization observed in the networks.

In practice, methods to capture the correct fibration partitioning over information-processing networks are, in general, based on what is called a negative iterative strategy~\cite{algDesign,PTarjan1985}. In this approach, an adequate initial partition is refined in each iteration with the goal to derive a better partitioning according to a predefined rule. Thus, each group of nodes in each iterative step is splitted into several, or none, smaller groups with respect to the whole current partition. This strategy defines the refinement paradigm~\cite{algDesign, refinetoolkit}. 
This refinement procedure is also used to find the so-called balanced equivalence relations~\cite{Aldis,Hasler2011,KameiCock} of directed networks, also known as balanced colorings. A balanced coloring over a network is a partition of the set of nodes obtained by defining an equivalence relation between the elements of the network~\cite{Stewart2006}. Thus, two nodes belonging to the same equivalence class, an element of the partition, can be permutated without changing the partition that is induced by the equivalence relation.
This relation, as defined in \cite{Stewart2006,fibration2019} regarding the study of robust synchronization patterns, is an equivalence relation between the input sets of the nodes, and it is defined by the input tree isomorphisms representative of the fibration symmetry \cite{Boldi2002,Boldi2006, fibration2019}.
Therefore, the nodes with the same color within a minimal balanced coloring also belong to the same fiber, where 'minimal' refers to the minimal number of classes that determine a balanced coloring (Equivalence is treated in section~VIIC of~\cite{IanGeneExpr2021}). Thus, the fibration partitioning over a directed network induced by symmetry fibrations is equivalent to the minimal balanced coloring over the same network. Throughout the rest of this work, we will use the terms "minimal balanced coloring" and "fibration partitioning" interchangeably. 

A few different algorithms have been proposed to obtain the minimal balanced coloring over directed networks. In this work we briefly describe the algorithms in~\cite{Aldis,Boldi2006} corresponding to the Aldis' algorithm and the Boldi-Vigna algorithm, respectively. We also present a more detailed description of the algorithm presented by Kamei and Cock as an extension of the algorithm proposed by Belykh and Hasler in~\cite{Hasler2011}, since this method is the one used in Morone \textit{et. al.}~\cite{fibration2019} to uncover the fibration partitioning over genetic networks and other real networks.
Although these algorithms are intuitive and easy to implement, they can be inefficient both in the time and memory complexities, meaning that they can be applied only for small to medium networks. Fortunately, because of the principled approach provided by the refinement paradigm and its relation with balanced colorings, we can make use of classical methods~\cite{refinetoolkit, modular,PTarjan1985,Tarjan1987} originally not designed for the identification of fibers to design very efficient algorithms that can be applied to obtain the minimal balanced coloring for large sparse networks. 

Here, we extend classical approaches to present an efficient algorithm capable of identifying the minimal balanced coloring for general information networks that also outperforms the alternative algorithms mentioned previously.
The algorithm built here is a modified version of the algorithm presented by Paige and Tarjan~\cite{Tarjan1987}, with runtime complexity of $\mathcal{O}(M\log N)$ and space complexity of $\mathcal{O}(M+N)$, where $M$ and $N$ are the number of edges and nodes in the network, respectively. This algorithm has the same runtime order than the algorithm presented by Cardon and Crochemore~\cite{Cardon1982}. However, compared with the algorithm of Cardon and Crochemore, the Paige-Tarjan(PT) algorithm has a simpler implementation and smaller prefactors, exhibiting a better performance to our problem. 
We note that both algorithms were not originally designed for the identification of fibers. Nevertheless, by introducing simple modifications we use the algorithm of Paige and Tarjan as a foundation of our method. We denote this method as Fast Fibration Partitioning (FFP) to constrast with the PT algorithm as here we apply the method specifically for the identification of fibers. We observe that this method outperforms the $\mathcal{O}(N^2\log N)$ time complexity and the $\mathcal{O}(N^2)$ memory complexity of the Kamei-Cock (KC) algorithm used in \cite{fibration2019}, providing a more efficient approach for further studies on cluster synchronization on complex networks. 


The paper is organized as follows. In Sec.~\ref{sec:coloring} we describe the balanced coloring algorithm used by Morone {\it et. al.} \cite{fibration2019} to contextualize the idea of refinement on fibration applications. In Sec.~\ref{sec:fastfiber} we explain how the general refinement paradigm can be modified to define the relationship between the input tree isomorphism and the algorithm we employ. 
In Sec.~\ref{sec:init} we give the details concerning the proper initialization and implementation of the algorithm. In Sec.~\ref{sec:alternative} we list the mechanisms of two additional approaches used for fiber identification. In Sec.~\ref{sec:comparison} we discuss all the algorithms introduced in order to compare their complexity. Through experiments on random networks, we also compare the runtime of the main current method for fibration and the method described in this work. Finally, in Sec.~\ref{sec:conclusion} we make our conclusions.

{


\section{\label{sec:coloring}Refinement for fibrations}

The refinement paradigm has been used widely in problems originated from computer science and discrete mathematics \cite{refinetoolkit,Tarjan1987}. It is also one of the major techniques used to develop efficient algorithms to solve the modular decomposition problem in graphs \cite{modular}. From state minimization in deterministic automata models \cite{hopcroftmin} to the ordering of boolean matrices and the lexicographically breadth-first search ordering of graphs \cite{Tarjan1987,LBFS2004,refinetoolkit}, the refinement procedure has been used to calculate the so-called balanced equivalence classes over a given set $V$ \cite{Tarjan1987,KameiCock}.
Beyond its applications, the concept of balanced equivalence classes has also shown significance because it is directly related to the patterns of dynamical processes occurring on complex networks \cite{SyncPattern}. 
These equivalence relations are constructed through successive refinement steps according to a disjoint set rule, which should guarantee that the final partition, obtained by a refinement algorithm, is a set of pairwise disjoint sets representing the balanced classes.

\subsection{The Belykh-Hasler algorithm}
\label{subsec:mbc}

In Morone \textit{et. al.} \cite{fibration2019}, the authors used a balanced coloring algorithm to find the correct fibration partitioning for several information networks. The original algorithm was first presented by Belykh and Hasler \cite{Hasler2011} and, even though they refer to the algorithm as a coloring algorithm, it differs from the classical coloring problems from graph theory. 
In the case of balanced coloring algorithms, adjacent nodes can have the same color and still be a proper coloring for the network. Precisely, in the Belykh-Hasler algorithm every node is identified by a unique color representing its class, where nodes having the same color belong to the same class of isomorphic input set nodes, meaning that these nodes are equivalent. Since each class does not overlap with any other, the balanced coloring obtained by the Belykh-Hasler algorithm represents a proper partitioning over the set of nodes $V$ of the network $G(V, E)$.

For the description of the algorithm used in \cite{fibration2019}, we define a coloring over the network as balanced if for every pair of nodes $v, w \in V$ belonging to the same class exists an isomorphism between their input sets, that is, there is a one-to-one relation between each element of the input set of $v$ with each element of the input set of $w$. 
If for every pair of nodes having the same color, they receive equivalent information via its incoming links, then the coloring is balanced and each color represents a label for one fiber of the network. For instance, in Fig.~\ref{fig:ex}(a) all nodes have different input trees, while nodes Y and Z have isomorphic input trees in Fig.~\ref{fig:ex}(b). Since there is a relation one-to-one between their input trees, both Y and Z synchronize their dynamical states. 
Thus, equivalent classes of a balanced coloring represents the proper fibration partitioning over a given network, implying that finding the minimal balanced coloring is equivalent to finding fibers over a minimal fibration.

Regarding the refinement paradigm, the Belykh-Hasler algorithm is based on the idea of choosing an initial coloring and refining this coloring until further refinement becomes impossible. Initial coloring or partitioning places all nodes inside the same equivalence class, and each iteration introduces an input driven refinement. 

To define an input driven refinement we introduce the Input Set Color Vector (ISCV) \cite{transistor2019}. The Input Set Color Vector of a node $v \in V$ is a $K$-dimensional vector, where $K$ is the number of colors in the network and the $j$-th entry of this vector counts how many nodes of color $j$ belongs to the input set of $v$.
A coloring is then balanced if, and only if, nodes of the same color have the same ISCVs. Moreover, a coloring is balanced and minimal if, and only if, nodes of the same color have the same ISCVs and nodes of different colors have different ISCVs.

In the algorithm, all nodes are initially assigned with the same color, representing the same equivalence class. Then, ISCVs for all nodes are calculated. Refined coloring needs to be balanced, therefore nodes of the same color need to have the same ISCVs. Refinement step is done by assigning all unique ISCVs with different colors and assigning refined colors to the nodes according to their ISCV. That is, the refined color of the node is defined by the color assigned to the ISCV of this node in the set of the unique ISCVs. If the coloring after the iteration is equivalent to the coloring before the iteration, the algorithm stops and we obtain a balanced coloring. An example of how the algorithm works is shown in Fig.~\ref{fig:minimalb}(a).



\begin{figure*}[t]
  \centering
  \includegraphics[scale=0.08]{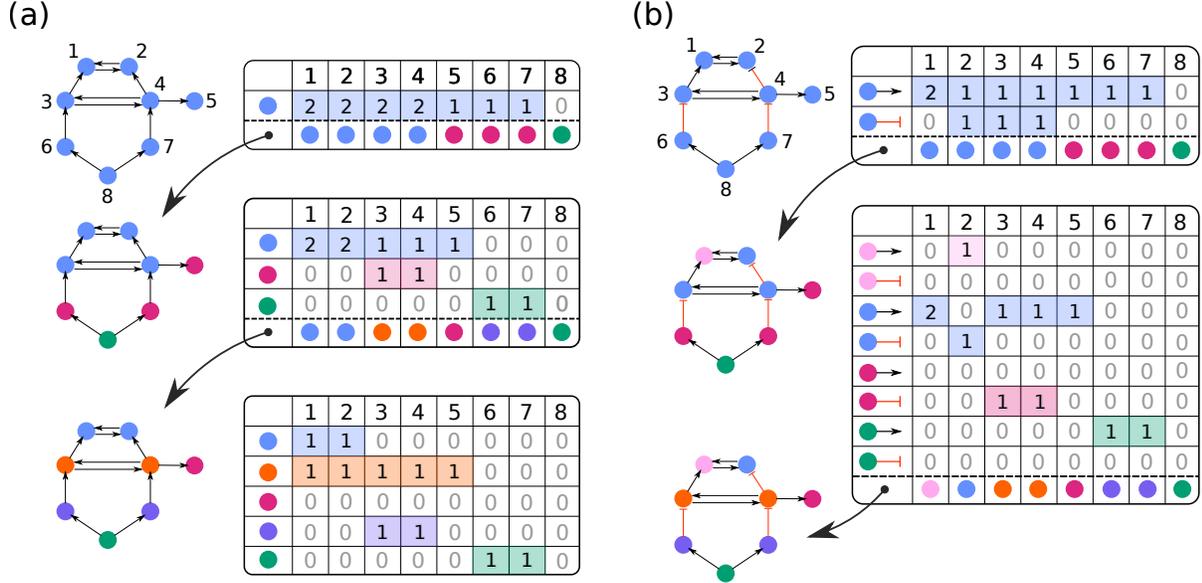}
  \caption{\textbf{Kamei-Cock algorithm}. Diagrams (a) and (b) show examples of how the Kamei-Cock algorithm works for two distinct networks. For both cases, we define in each iteration the matrix of ISCVs to verify the balancing of the current coloring. In the process, every time the number of colors increases, the matrix increases as well. This process continues until the matrix is balanced and no new colors are necessary.  In (a), we consider a network containing 8 nodes. Initially, all nodes are assigned to a light blue color. The ISCVs have only one element (one color) and its value is shown at the top left table. Here, all ISCVs are showed together as a matrix. Node $1$ receives inputs from nodes $2$ and $3$, while node $2$ receives input from $1$ and $4$. Thus, both nodes have two inputs of color blue. We note that only nodes $3$ and $4$ also have two inputs from color blue. From this, nodes $1$ to $4$ will belong to the same class in the next iteration. In total, there are three unique vectors. Each one of these unique vectors is assigned to a new color and a new partition is created. This time, the matrix of ISCVs grows to dimension $8\times3$, reflecting the increasing on the number of colors. Calculating again the ISCVs, we see that the coloring is not balanced, such that a new refinement is possible. At last, considering the bottom left matrix, we finally arrive to the final balanced coloring, with no new refinement possible, providing the fibers of the given network. For multidimensional edges, for each new color added during the refinement, the ISCV for each node is increased by $K_{type}$, where $K_{type}$ is the number of edge types. As shown in (b), the presence of new edge's types in the same network of (a) can generate a different balanced coloring.} 
  \label{fig:minimalb}
\end{figure*}

\subsection{Balanced coloring for multiple edge types}

The approach presented in the section~\ref{subsec:mbc} was extended in 2013 by Kamei \& Cock \cite{KameiCock} to be applied for directed graphs with multiple types of edges. This algorithm can be applied to a genetic network with activator and repressor edges, for instance. In this case, instead of considering only the color of the node in the input set both node's color and type of the edge are considered. All the steps taken by the algorithm remain the same, but the size of a ISCV is now equal to the number of colors in the graph multiplied by the number of edge types. 
For instance, let us consider the same graph as before, but edges now have different types, {as shown in Fig.~\ref{fig:minimalb}(b). This time, initial ISCVs are two-dimensional (see the top matrix in Fig.~\ref{fig:minimalb}(b)) to account for the two types of edges. From this, the same process as before is repeated until a balanced coloring is obtained and no further refinement is possible. By introducing different types of edges, we obtain a different fibration partitioning compared to Fig.~\ref{fig:minimalb}(a). Finally, the refinement process for the Kamei-Cock (KC) algorithm is defined by the following steps: 

\begin{enumerate}
  \item Define the initial coloring $\bar{C}_0$, where each node $v \in V$ has a color label $\mathcal{\ell}(v)$ associated, and calculate the initial number of colors $K$;
  \item For each node $v$, define its ISCV$(v)$;
  \item Find all the unique ISCVs, assign to each one a different color label, and calculate the new number of different colors $K'$;
  \item The color $\mathcal{\ell}(v)$ of each node $v$ is set as equal to its ISCV: $\mathcal{\ell}(v) \leftarrow \mathcal{\ell}(ISCV(v))$;
  \item If set $K \neq K'$, then $K \leftarrow K'$ and go to step 2. Otherwise, the algorithm stops.
\end{enumerate}
\label{alg:general-algo}

Furthermore, there are three important points to be noted for this algorithm: 1) The number of dimensions of the ISCV grows quickly; 2) The matrix of ISCVs is very sparse, as we can see in the examples shown in Fig.~\ref{fig:minimalb}; 3) Initial coloring can be assigned to certain nodes. These details are very important specially when the task is to identify fibers in large networks. Therefore, it is easy to grasp the limitations of the method concerning its storage efficiency. For the networks treated in \cite{fibration2019}, the KC algorithm does not show any major issue since the authors were only interested with relatively small networks. However, since information network data are continually growing and getting more complex \cite{bigBioData,MarxBigData}, it is pertinent that a method to identify fibrations on large networks should be as efficient as possible. Based on that, we describe in the following section a different algorithmic approach, still based on the refinement partitioning paradigm, capable to handle efficiently both with the storage and the runtime in general applications.

\section{Optimal identification of network fibers}
\label{sec:fastfiber}

According to the concept of graph fibrations \cite{Boldi2002}, all nodes inside the same set must receive equivalent information from all the other sets, meaning that the desired partition splits the network into several groups, called classes, each one containing nodes with isomorphic input trees. To obtain that, we treat our problem as the same as finding the coarsest relational refinement partitioning of a set of elements that have a binary relation between them~\cite{Tarjan1987}. Since a network is completely defined by a set of node elements and a set of edges that comprehend all the binary relations, this approach can be used. Moreover, because this paradigm should obey the fibration rules, in what follows: ({\it i}) we show how a refinement method can be constructed to identify the fibration isomorphisms, and ({\it ii}) we present the algorithm used to find the coarsest partition in this context and show the implementation of the proposed method.

\subsection{Partition refinement paradigm for input tree isomorphism}
\label{ssec:fiber-refinement}

Considering the application for a directed network $G(V, E)$, defined by $N = |V|$ nodes connected by $M = |E|$ edges, we can define a network or graph partition $\bar{P}$ over $V$ as a set of pairwise disjoints subsets $\raisebox{2pt}{$\chi$}^j \subseteq V$ whose union is all $V$, that is
\begin{equation}
  \bar{P} = \{ \raisebox{2pt}{$\chi$}^j \subseteq V | \raisebox{2pt}{$\chi$}^i \cap \raisebox{2pt}{$\chi$}^j = \emptyset, i \neq j \}
\end{equation}
and
\begin{equation}
	V = \bigcup_{j} \raisebox{2pt}{$\chi$}^j,
\end{equation}
where $\raisebox{2pt}{$\chi$}^{j}$ are the elements of the partition $\bar{P}$, called classes. By taking an additional graph partition $\bar{Q}$ with the property that each one of its classes are contained in the classes of $\bar{P}$, we say that $\bar{Q}$ is a refinement of $\bar{P}$, or $\bar{Q} \preceq \bar{P}$ \cite{refinetoolkit}. Moreover, considering a proper refinement, a partition $\bar{P}$ is said to be stable if it is stable with respect to all its classes $\raisebox{2pt}{$\chi$}^j$. In classical applications, for a class $\raisebox{2pt}{$\chi$} \in \bar{P}$, we say that $\raisebox{2pt}{$\chi$}$ is stable with respect to a set $S$ if either all elements of $\raisebox{2pt}{$\chi$}$ connects with an element of $S$ or none element of $\raisebox{2pt}{$\chi$}$ points to any element of $S$. Thus, the goal of a refinement partitioning algorithm is to find the coarsest (minimal number of classes) stable partition over the set $V$, given a relation $E \subseteq V \times V$. As can be noted, this can be easily extended to graph problems, where the coarsest graph partitioning problem is that of finding, for a given set of directed edges $E$ and an initial partition $\bar{P}$ over $V$, the minimal number of disjoint classes that are subsets of $V$ forming a stable refinement of $\bar{P}$.

The stability of partitions is the most important concept for the proper identification of equivalence classes and to guarantee the correctness of the algorithm and its termination proof. In the work of Paige and Tarjan \cite{Tarjan1987}, the authors define the stability properties through the definition of a function $s(S, \bar{Q})$ responsible for splitting the classes of the input partition $\bar{Q}$ into stable classes with respect to a set $S$, also called \textit{pivot set}. The function $s$ defines the \textit{refine} operation, responsible for the refinement of an input partition with respect to a pivot set $S$, replacing this current unstable partition for a new one that is stable with respect to $S$. If $s(S, \bar{Q})$ behaves as an identity function for $\bar{Q}$, meaning $\bar{Q} = s(S, \bar{Q})$, then the partition $\bar{Q}$ is already stable with respect to $S$. The following two properties described in \cite{Tarjan1987} are the most relevant to our algorithm: the stability inheritance by refinement and the stability inheritance under union. Precisely, any refinement of $\bar{P}$, where $\bar{P}$ is stable with respect to $S$, is also stable to $S$, and a partition stable with respect to two different pivot sets $S$ and $S'$ is also stable with respect to their union $S \cup S'$.

The main advantage of the definitions presented is that the stability rule can be modified to represent a more suitable disjoint rule and still preserve the stability inheritances mentioned. With this, we can define a new refinement stability criterion that permits the direct relationship between the refinement partitioning paradigm with the notion of isomorphism between input trees in information networks. Thus, we can use the concept of graph fibrations to determine a disjoint rule that allows the refinement algorithm to find the coarsest graph partitioning where elements inside the same class \raisebox{2pt}{$\chi$} of the final partition all have isomorphic input trees. This way, by a simple modification, we introduce the concept of \textbf{input set stability} and its corresponding splitting function $s_{input}$ for the \textit{refine} operation.

In order to identify the group of nodes with isomorphic input trees, we require that the partition should be \textbf{input set stable} during each refinement with respect to the pivot sets. In that manner, a graph partition $\bar{P}$ over the network $G(V, E)$ is input set stable with respect to $S \subseteq V$ if, for all the classes $\raisebox{2pt}{$\chi$} \in \bar{P}$, the following equality is satisfied for all the elements $v$, $w \in \raisebox{2pt}{$\chi$}$ and for all the types $k$ of directed edges:
\begin{equation} \label{eq:input-set}
	| E_{k}^{-1}(\{v\}) \cap S | = | E_{k}^{-1}(\{w\}) \cap S |,
\end{equation}
where $E_k(\{v\})$ and $E_k^{-1}(\{v\})$ represent, respectively, the nodes that directly receives information of type $k$ from $v$, and the nodes that sends information via a directed edge of type $k$ to $v$. We stress that both $E_k(\{v\})$ and $E_k^{-1}(\{v\})$ may contain repeated elements. From that, an input set stable graph partition implies that each class receives equivalent information from all the other classes, including from itself. From this, it is straightforward to check that the inheritance by refinement and the inheritance by union of classes is also valid for $s_{input}$, where the rule given by the Eq.~(\ref{eq:input-set}) guarantees the partitioning in disjoint sets.

After we find the coarsest stable graph partition $\bar{P}$ following the input set stability rule there will not be any pivot set $S \subseteq V$ and $S \not\subset \raisebox{2pt}{$\chi$}$ for which the partition $\bar{P}$ is not input-set stable, meaning that for any class $\raisebox{2pt}{$\chi$} \in \bar{P}$ we have $\bar{P} = s_{input}(\raisebox{2pt}{$\chi$}, \bar{P})$. 
As a consequence, any two nodes belonging to the same class receive equivalent information from the rest of network, including from their own class, implying that they have the same input set. Fortunately, if two nodes in an input set stable network partitioning have the same input set we can use the results presented by Norris~\cite{norris1995} to show that these nodes have isomorphic input trees. 
According to Norris~\cite{norris1995}, for a directed network with $N$ nodes, an isomorphism for any depth $k$ between the input trees of two nodes can be guaranteed as long as these trees are isomorphic up to the $N-1$ depth, where the depth $k$ of a tree is the set of nodes with $k$ edges between them and the root node.
Therefore, we can state that the coarsest input set stable graph partition obtained by the refinement procedure using $s_{input}$ is the union of pairwise disjoint sets, where each one of these corresponds to a set of nodes with isomorphic input trees. This result shows a natural mapping between the refinement partitioning paradigm and the identification of fibers in information-processing networks.

\subsection{Fast fibration partitioning}

As we have discussed in the last section, to refine a partition it is necessary a pivot set $S$ to split the classes of the current partition into different classes, each one input set stable with respect to $S$. In order to do that, one should choose an appropriate list of pivot sets.
For instance, if we consider a partition $\bar{P}$, to show that it represents the correct partitioning in groups of isomorphic input tree nodes, it is necessary to verify that $\bar{P}$ is input set stable with respect to each one of its classes. By doing so, it is possible to design an algorithm capable of determining the fibration partitioning of a network by a general straightforward procedure: starting with an initial network partition, refine the current partition with respect to each one of its classes until there is no more unstable class. 
If applying $s_{input}$ for every class of the current partition leads to no further refinement, then the current partition is input-set stable and correspond to the correct minimal balanced coloring of the network. However, even though this procedure gives the right answer in a reasonable runtime and it represents the essence of a refinement algorithm, it also gives a high number of unnecessary and repeated operations. Therefore, it is important to choose the appropriate pivot sets during the refinement in order to design an efficient method for the partitioning.

The algorithm we build here is a modified version of the algorithm introduced by Paige and Tarjan~\cite{Tarjan1987}, with runtime complexity of $\mathcal{O}(M\log N)$. Although this algorithm was designed for a different problem, namely the relational coarsest partition problem, we emphasize that this problem is easily mapped to the identification of fibers on graphs, as pointed out in section~\ref{ssec:fiber-refinement} and in references \cite{Boldi2002,Boldi2006}.
By introducing the input set stability definition for graphs, we use the algorithm of Paige and Tarjan as a foundation of our method. Compared to the similar algorithm of Cardon and Crochemore \cite{Cardon1982}, the Paige-Tarjan algorithm has a simpler implementation and smaller runtime prefactors, since it avoids irrelevant pivot set selection, exhibiting a better performance to our problem.

For the identification of an input set stable graph partition, we can benefit from the input set stability properties to construct a refinement algorithm that should achieve, through a finite number of steps, the minimal input set stable partition originating from an initial trivial network partitioning. A refinement step should have the effect to refine the current partition of $V$, unstable with respect to a pivot set $S \subseteq V$, by replacing it for a new partition, now stable for $S$. For this, we use the splitting function $s_{input}(S, \bar{P})$, which receives, as input, the current partition $\bar{P}$ over the network and the pivot set $S$, returning as output a new stable partition with respect to $S$. As pointed out, $s_{input}$ benefits from two stability properties: the inheritance by refinement and the inheritance by union of sets. Because these inheritances, a pivot $S$ should be used only once by $s_{input}$, which guarantees that the partition, after all refinement steps, will preserve the stability with respect to $S$ and to the union $\cup_i S_i$ of all pivot sets $S_i$ already used. From these properties, the core of the \textit{refine} operation can be stated as the three following steps:
\begin{enumerate}
	\item Choose a pivot set $S$ in which the current graph partition $\bar{P}$ is input set unstable;
	\item  Identify all the classes $\raisebox{2pt}{$\chi$}_{/S} \in \bar{P}$ that are input set unstable with respect to $S$ and split each one of these into stable classes $\raisebox{2pt}{$\chi$}_S^{j} \subset \raisebox{2pt}{$\chi$}_{/S}$ with respect to $S$;
	\item Replace $\bar{P}$ by $\bar{P}_S = s_{input}(S, \bar{P})$, where for $\bar{P}_S$ each $\raisebox{2pt}{$\chi$}_{/S}$ is replaced by its splitted classes $\raisebox{2pt}{$\chi$}_S^{j}$.
\end{enumerate}

Through these steps, the \textit{refine} operation can be executed with a time complexity of the order $\mathcal{O}(|S| + \sum_{v \in S} |E(\{v\})| )$.
To obtain this time complexity we should first note that the only possible input set unstable classes $\raisebox{2pt}{$\chi$}_{/S}$ from $\bar{P}$ are the ones who receives input from the nodes of $S$. Therefore, we need to iterate over all $|S|$ nodes $v \in S$ and then iterate over all the $|E(\{v\})|$ outgoing edges of each $v$ to properly define all unstable classes $\raisebox{2pt}{$\chi$}_{/S}$, which are the ones that contain nodes where the condition from Eq.(\ref{eq:input-set}) is not satisfied. Thus, the \textit{refine} operation is linear either with the number $\sum_{v \in S} |E(\{v\})|$ of outgoing edges from $S$ or with the number of nodes $|S|$ depending on which one is larger.
This ensures the \textit{refine} operation as optimal, since we only need to identify the outgoing information leaving each element of $S$ and arriving to the possible input set unstable classes. Since the finest partitioning possible is the one in which every node is itself a class, and because the number of classes increases after each refinement step, the refinement by pivot sets may be executed at most $N - 1$ times.
Furthermore, because of the splitting dynamics of the algorithm, each node $v \in V$ is in at most $\log N$ different pivot sets $S$, as also described in~\cite{Tarjan1987}. Therefore, using the time complexity of the \textit{refine} operation for every node, used as a part of a pivot set at most $\log N$ times, we obtain the $\mathcal{O}(M\log N)$ runtime of the algorithm. 

 

Through this, given a pivot set $S$ and a given input set unstable network partition $\bar{P}$, the classes $\raisebox{2pt}{$\chi$}_{/S} \in \bar{P}$ that are unstable with respect to $S$ can be splitted into several disjoint classes $\raisebox{2pt}{$\chi$}_S^{j}$ input set stable with respect to $S$. Considering the general situation for multiplex networks containing $n$ types of edge, we can denote the splitted classes as $\raisebox{2pt}{$\chi$}_S^{j_1j_2...j_{n}}$ related with their unique tuple $(j_1,j_2,...,j_n)_S$ regarding their stability with respect to $S$. Then, each splitted class must obey, for $k$ from $1$ to $n$, the property defined by
\begin{equation} \label{eq:gen_split}
	\raisebox{2pt}{$\chi$}_S^{j_1j_2...j_{n}} = \{ v \in \raisebox{2pt}{$\chi$}_{/S} : | E_{k}^{-1}(\{v\}) \cap S | = j_k \},
\end{equation}
where the number of splitted classes must be larger than one. Among all splitted classes obtained from $\raisebox{2pt}{$\chi$}_{/S}$, the largest one can be excluded as pivot in the following steps of the algorithm. This is a consequence of the disjoint nature of the refinement process and its stability inheritance property. For instance, if we refine a partition with respect to a pivot set and that pivot is responsible for splitting one unstable class $\raisebox{2pt}{$\chi$}$ into two stable classes, $\raisebox{2pt}{$\chi$}_1$ and $\raisebox{2pt}{$\chi$}_2 = \raisebox{2pt}{$\chi$} - \raisebox{2pt}{$\chi$}_1$, then it is necessary to check only the partition stability with respect to one of these splitted sets. The complement set is just a redundant pivot, due to the fact that $\raisebox{2pt}{$\chi$}_1 \cap \raisebox{2pt}{$\chi$}_2 = \emptyset$ and that the refined partition is input set stable with respect to $\raisebox{2pt}{$\chi$} = \raisebox{2pt}{$\chi$}_1 \cup \raisebox{2pt}{$\chi$}_2$. This ensures that none repeated, redundant or union of repeated sets is used during any step of the algorithm.

At this point, we can finally state the complete algorithm to identify the correct partition $\bar{F}$ of a directed network $G(V, E)$, representing a balanced coloring. The first step is to define the initial partitioning $\bar{P_0}$ which requires some preprocessing. We let the details of this step for the next section. Here we can consider the simpler case, where $\bar{P_0} = \{V\}$. We also define $L$ as a queue data structure to store the pivot sets during the execution of the algorithm. From this, the algorithm, which we denote as fast fibration partitioning (FFP), is given by the following steps: 
\begin{enumerate}
    \item Define the initial partition $\bar{P}_0$, push all its classes to a queue $L$, and set $\bar{P} = \bar{P}_0$;
    \item Remove from $L$ its first pivot set $S$;
    \item Replace $\bar{P}$ by $s_{input}(S, \bar{P})$;
    \item Whenever a class $\raisebox{2pt}{$\chi$} \in \bar{P}$ splits into two or more nonempty classes, push all these to the back of $L$, except the largest one;
    \item If $L$ is not empty, go back to step 2. Otherwise, the algorithm stops and the final partition $\bar{F}$ represents the minimal balanced coloring.
\end{enumerate}
At the end, the final partition $\bar{P} = \bar{F}$ represents the coarsest input set stable partition, or minimal balanced coloring, of $G(V, E)$ and each class of $\bar{F}$ is a fiber.


The final algorithm has time and space requirements of order, respectively, $\mathcal{O}(M\log N)$ and $\mathcal{O}(M+N)$, which allows its application for large networks. It has a simple implementation, where the refinement process, pictured in Fig.~\ref{fig:refine}, represents the most complex part of the algorithm.

\section{Initialization and implementation details}
\label{sec:init}

As pointed out, to correctly implement the refinement algorithm it is very important that the initial partition, represented by $\bar{P_0}$, satisfies determined conditions to ensure that the nodes will not be placed inside wrong classes of the final partition $\bar{F}$. For this, given the network $G(V, E)$ and before any refinement, we can define specific groups of nodes based on their input from other nodes. These groups are related with the network $G_{scc}(V_{scc}, E_{scc})$, obtained by the partitioning of $G$ into strongly connected components (SCC). In this way, each element of $V_{scc}$ represents the set of nodes that are reachable from all the other nodes in this same element via a directed path. For convenience, we define isolated nodes in $V_{scc}$ as components themselves. From this, to initialize the partition $\bar{P_0}$, we assign to each component $v' \in V_{scc}$ a class label according to the condition whether $v'$ receives or not input from other components $w' \in V_{scc}$, where $v' \neq w'$. This step zero is necessary since two different SCCs that do not receive any input from other elements cannot have correlated dynamics.

\begin{figure}[t]
  \centering
  \includegraphics[scale=0.412]{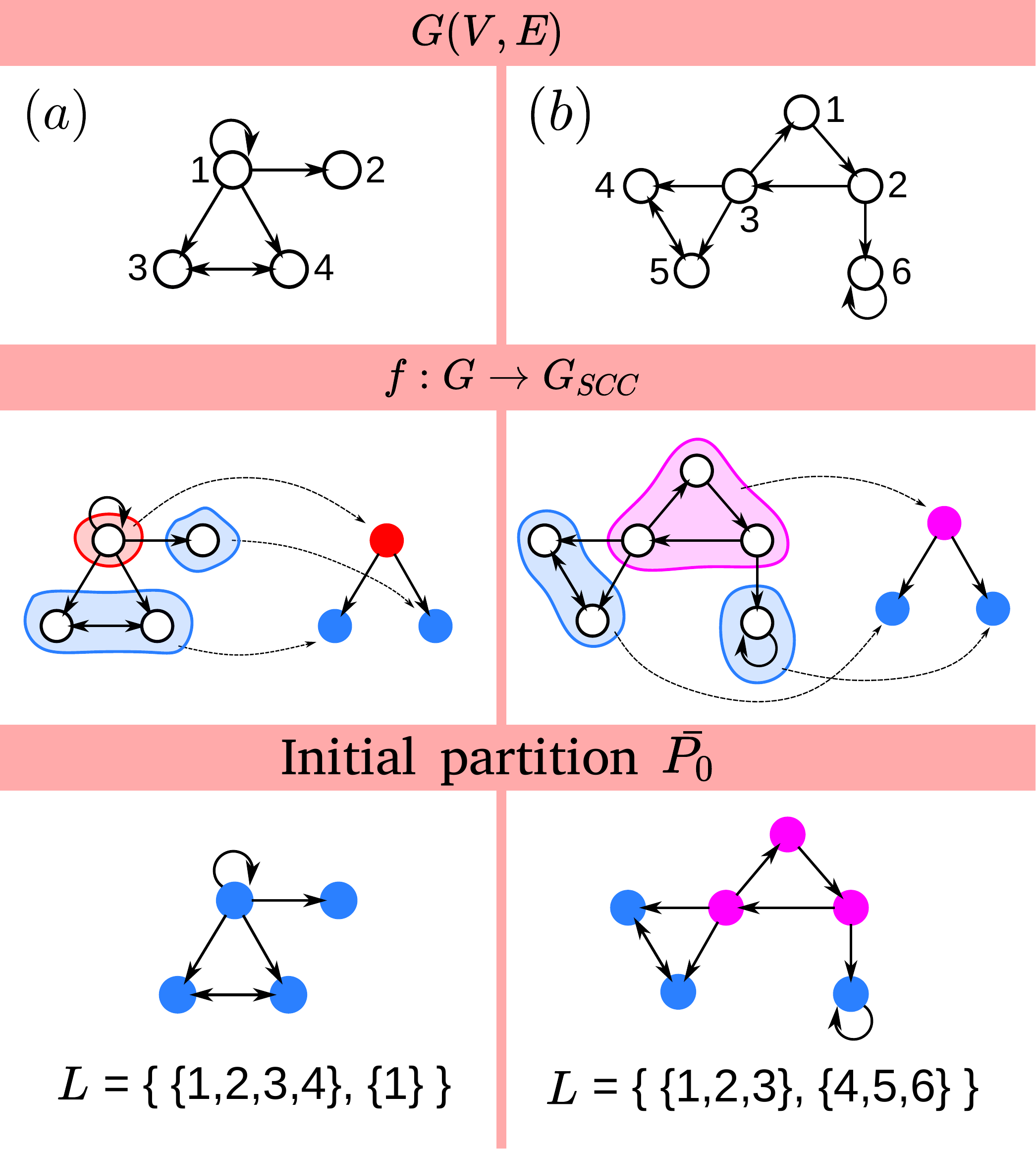}
  \caption{\textbf{Definition of the initial partition $\bar{P_0}$.} Figures (a) and (b) display two small network examples where we need to take into account different conditions to define correctly the initial partition $\bar{P_0}$. Starting from a given network $G(V,E)$ we first obtain all the strongly connected components (SCC) present in the network to define $G_{SCC}$. 
  After that, we classify each node of $G_{SCC}$ according to three different conditions. In both (a) and (b) we observe the presence of SCCs that receive input from other components, and they are highlighted with the color blue. All nodes of $G$ belonging to blue nodes of $G_{SCC}$ must be placed into the same class $\raisebox{2pt}{$\chi$}_0$ of $\bar{P_0}$. Specifically in (b) we note the existence of a strong component that do not receive input from any other different component. This component is highlighted with color pink. For each pink element $v' \in V_{SCC}$ we define a new class $\raisebox{2pt}{$\chi$}_{v'}$ in $\bar{P_0}$ containing the nodes inside this element.
  Moreover, in (a) we highlight by color red a third type of component. These SCCs are comprised by single nodes that only receive inputs from themselves. The nodes belonging to these components are also placed in the same class of blue nodes. However, we also push these single nodes as classes $\raisebox{2pt}{$\chi$}_{self} = \{ v_{self} \}$ into queue $L$, where here we denote as $v_{self}$ any isolated self-loop node from $G$.}
  \label{fig:initialization}
\end{figure}

We define three types of SCC that must be separated before the application of the refinement process. The first type, and more trivial component, is any SCC that receives external input from any other SCC (blue nodes in Fig.~\ref{fig:initialization}). All nodes belonging to these components are put into the same unique class $\raisebox{2pt}{$\chi$}_0$ at the beginning of the algorithm. 
The second type defines the opposite scenario, that is, any SCC that does not receive any input from the rest of the network (purple nodes in Fig.~\ref{fig:initialization}). Each component, represented by $v' \in V_{scc}$ of this type defines one different class $\raisebox{2pt}{$\chi$}_{v'}$ to be put in the initial partition. Besides bigger SCCs with no inputs, this second type allows for isolated nodes without incoming edges in $G$. 
However, the existence of these isolated components are not, in general, feasible in some real applications, like genetic regulatory networks. In this case, for every gene there should be at least one transcription factor responsible for its regulation, implying that all nodes in these types of networks have input, even if only from themselves. Of course, that is not necessarily true for general information networks, meaning that the second type definition remains valid even for these special cases.

\begin{figure*}[t]
  \includegraphics[scale=0.075]{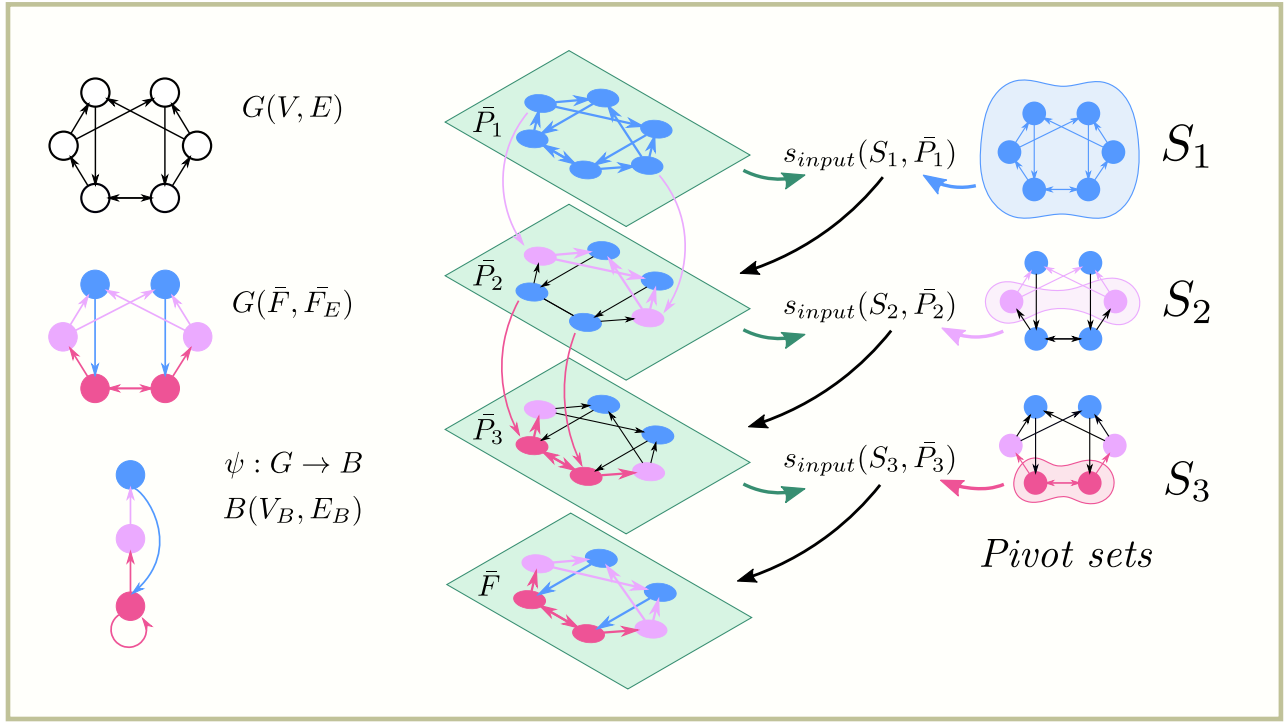}
  \caption{\textbf{Refinement mechanism of the splitting function} \bm{$s_{input}$}. Here we consider the directed network $G(V, E)$ shown at the left. We start initializing the algorithm by defining the first pivot set $S_1$ as the initial partition $\bar{P_1}$, containing all nodes of the network. The initial state has $\bar{P_1}$ as partition and $L = \{ S_1 \}$ as the pivot set queue structure. Following the refinement procedure, we select $S_1$ as the first pivot set, and then refine the current partition obtaining $\bar{P_2} = s_{input}(S_1, P_1)$, which splits the single blue class into two disjoint classes input set stable with respect to $S_1$: a upgraded reduced blue class and the new light-pink class. Since the blue class is the largest splitted block (containing four nodes against two from the pink class), we insert the pink class at the back of $L$, defining it as the pivot set $S_2$. Because $L = \{S_2\}$, we repeat the process this time using $\bar{P_2}$ and $S_2$ as input for $s_{input}$, which returns a new partition $\bar{P_3} = s_{input}(S_2, \bar{P_2})$. The blue class is splitted into two equal size classes and the new one is used as the pivot set $S_3$, right after be placed at the back of $L$. However, $\bar{P_3}$ is input set stable with respect to $S_3$, implying that none class of $\bar{P_3}$ is splitted. For this reason, $L$ is now empty and the algorithm stops. The final partition $\bar{F}$ represents the minimal balanced coloring over $G(V, E)$. As a consequence of the partition $\bar{F}$, the set of directed edges is also defined as a partition $\bar{F_E}$, which distinguishes the information coming from different fibers. Finally, from this partitioning, we can define the new network $B(V_B, E_B)$, representing the base network that resulted from the fibration morphism $\psi: G \rightarrow B$.}
  \label{fig:refine}
\end{figure*}


A more special case, very common in genetic networks in bacteria, is the existence of self-loops. To deal with these situations, we define a third type of SCC as a special case of the second type described above. This one is formed by components with a single self-loop node receiving input only from itself (red nodes in Fig.~\ref{fig:initialization}). 
Differing from the second type, this node is inserted into the $\raisebox{2pt}{$\chi$}_0$ in the same fashion as the nodes belonging to the first type components. However, now we define a different class $\raisebox{2pt}{$\chi$}_{self}$ containing only the self-loop node and push this class into the pivot set queue $L$. This rule is a consequence of the potential of these special nodes to split specific configuration of classes. 
Even when these configurations are input set stable with respect to all other classes in the partition, they can be further refined by their internal isolated self-loop nodes. Yet, nodes from $\raisebox{2pt}{$\chi$}_{self}$ cannot be initialized as different classes in the partition. That is why we only place these classes into the queue $L$. Regarding the KC algorithm presented in the section~\ref{sec:coloring}, we stress that only the classes $\raisebox{2pt}{$\chi$}_0$ and $\raisebox{2pt}{$\chi$}_{v'}$ (including self-loop nodes) are useful for its initialization. The initialization step can be described by the following steps: (i) define the partitioning $G_{scc}$ into elements as SCC. (ii) Then, classify each component exactly as previously explained. 
Put all nodes belonging to the first type components into the class $\raisebox{2pt}{$\chi$}_0$, create for each second type component $v'_j \in V_{scc}$ a new class $\raisebox{2pt}{$\chi$}_{v'}^j$ and, after this, push all these classes into the pivot set queue $L$. (iii) At last, consider all the isolated self-loop nodes $v'_k \in V_{scc}$, put them inside $\raisebox{2pt}{$\chi$}_0$, create the classes $\raisebox{2pt}{$\chi$}_{self}^k$ for each one, to finally push all $\raisebox{2pt}{$\chi$}_{self}^k$ into the queue $L$. After these steps, the initial state of the algorithm will be defined by $\bar{P_0}$ and $L$ as follows:
\begin{equation}
  \begin{cases}
    \bar{P_0} & = \{ \raisebox{2pt}{$\chi$}_0 ,\ \raisebox{2pt}{$\chi$}_{v'}^j \ \}, \\
    L & = \{ \raisebox{2pt}{$\chi$}_0 ,\ \raisebox{2pt}{$\chi$}_{v'}^j ,\ \raisebox{2pt}{$\chi$}_{self}^{k} \},
  \end{cases}
\end{equation}
allowing the proper application of the FFP algorithm.

For the implementation, we can define the partition $\bar{P}$ holding all the classes by a doubly linked list, which allows fast deletion of classes in constant time $\mathcal{O}(1)$ as long as we have the memory address of the class during the procedure. Alternatively, we can define $\bar{P}$ as a simple indexed list containing classes. If we choose to not perform the deletion of classes, each class is a doubly linked list. Each component of the linked list is a node belonging to the class. This structure allows that a class $\raisebox{2pt}{$\chi$}$ does not have to be deleted during the splitting process, but only reduced by removing its nodes. This way, instead of deleting classes we only need to create new ones to store the nodes removed from other classes. Moreover, it is necessary that each node at each iteration has a pointer to its class, or simply to the index in the list $\bar{P}$ where its class is located. This pointer is updated whenever a class is splitted. Since the splitting condition reduces to checking Eq.~(\ref{eq:gen_split}), and $L$ can be implemented by a simple queue, the data structures described here are the main ones for an efficient implementation of the algorithm. 

\begin{figure*}[t]
  \centering
  \includegraphics[scale=0.06]{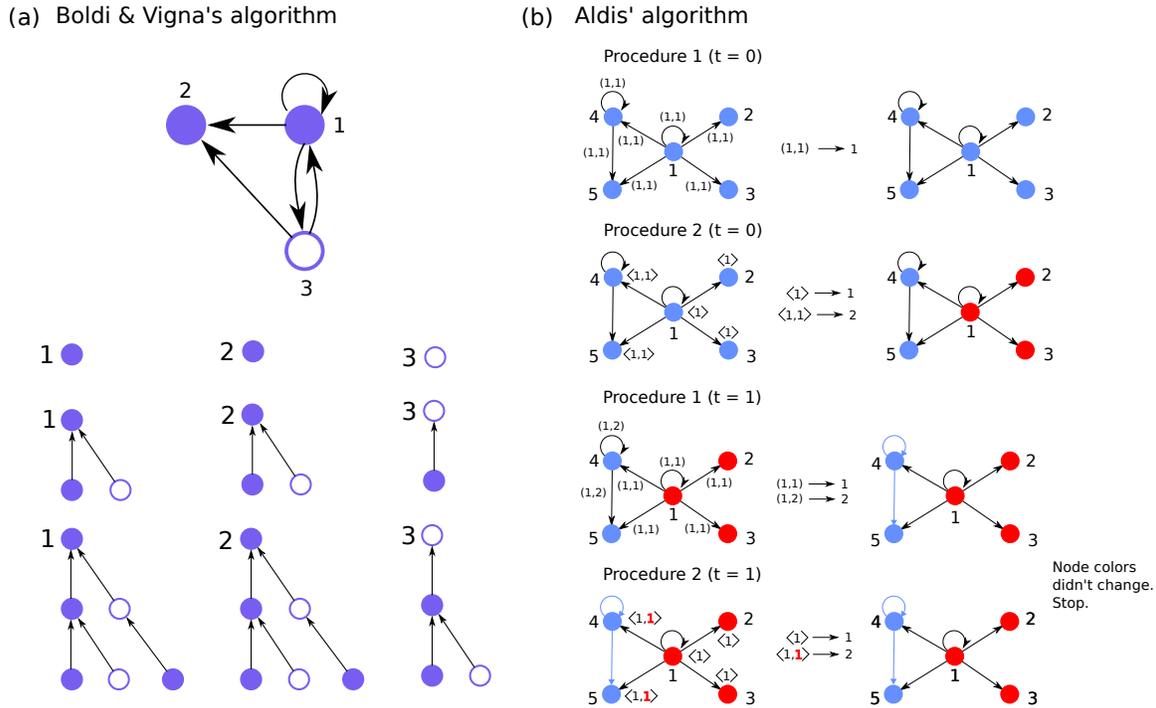}
  \caption{\textbf{Alternative algorithms to find fibers}. In (a) we see a Fibonacci circuit, introduced in \cite{fibration2019}. Following the approach of Boldi and Vigna \cite{Boldi2002}, at the initial step all 3 nodes have the same input tree of depth 0, so initial partition puts all nodes in the same class. The second step shows that node $3$ is already non-isomorphic to $1$ and $2$, however we need to go further $N-1 = 2$ levels to assign the final isomorphism between input trees. At the final level, the algorithm assigns the isomorphism between the input trees of $1$ and $2$, defining the Fibonacci fiber. Finally, (b) exemplifies the Aldis'\cite{Aldis} algorithm for the case detailed in section \ref{sec:alternative}.} 
  \label{fig:boldi_aldis}
\end{figure*}

The reader can refer to the link \url{https://github.com/makselab/} to access the Julia implementation of the fast fibration partitioning algorithm as described in this work. R library for finding fibration partitioning applying the algorithm developed in \cite{fibration2019} along with functionality for fiber building blocks construction, classification and significance detection are available at \url{https://github.com/makselab/fibrationSymmetries}. All codes are written in high performance languages and further instructions and documentation for their proper use can be found in the referred links.

\section{Alternative methods}
\label{sec:alternative}

As already mentioned, fibers are equivalent to the classes of a minimal balanced coloring over a network, which can be obtained by the methods presented so far. In what follows we present two different approaches previously developed to find balanced colorings that can be used to identify fibration symmetric circuits, and then we compare the runtime and space complexities of the algorithms with ours.

\subsection{Boldi \& Vigna's algorithm}

The first approach was proposed in 2002 by Boldi \& Vigna in their seminal work of graph fibrations \cite{Boldi2002}. Presented originally as a theorem, the idea is to construct the input tree for each node in order to verify the isomorphism between the nodes' input tree, and use the theorem by Norris \cite{norris1995} as the stop criterion. 
The initial partition is defined as a single class containing all the nodes, and the refinement operation is based on the search for isomorphism through the first layer until the ($N-1$)-th layer of the input trees. Two nodes are said to be equivalent under $\simeq_k$ if their input trees are isomorphic up to level $k$. Since a node is always isomorphic to another node, therefore all nodes are equivalent under $\simeq_0$. 
Moreover, an equivalence relation states that $v \simeq_{k+1} w$ $\iff$ $v\simeq_{k}w$ and then there exists a bijection $\psi$ from the input set of $k$-th layer of the input tree of $v$ to the input set of $k$-th layer of the input tree of $w$, such that it satisfies the isomorphism condition, \textit{i.e.}, the image of the edge connects the image of the source and the image of the target of this edge. 

In Fig.~\ref{fig:boldi_aldis}(a) we show an example for a Fibonacci circuit introduced in \cite{fibration2019}. For this type of circuit, the number of nodes in a $k$-th level is described by a Fibonacci sequence and, to obtain a new partition, we need to check the isomorphism condition for the number of elements in all levels up to $N - 1$. In \cite{Boldi2002}, the authors propose an algorithm to check the isomorphism between input trees with total runtime of order $\mathcal{O}(N^2d\log N)$, where $d$ is the maximum in-degree in the network, and memory of order $\mathcal{O}(N\log N)$. However, this runtime can still be very restrictive for large networks, not being suitable for general applications.





\subsection{Aldis' algorithm}

An alternative approach for the Kamei-Cock algorithm was proposed by Aldis \cite{Aldis} in 2008. According to his algorithm, all nodes are placed inside the same equivalence class and further partitions are obtained by a refinement process based on the type of the edge and the source of this same edge, similar to the input driven refinement introduced in Sec.~\ref{sec:coloring}. 
Aldis' approach consists of two main procedures, both of which are implemented in every iteration. We use the notation $n_t(v), n_t(e) \in \mathbb{N}$ to describe the color of the node $v$ or the edge $e$ in the iteration $t$. Once the algorithm stops, the balanced coloring is obtained.

In the first procedure, each edge is associated with a pair of the edge color and the edge source color $( n_t(e), n_t(s(e)) )$. Then, we list all unique pairs at the current iteration. Each unique pair is associated with a new edge color and the edge coloring of the network is updated. If there are no new colors after the update at this iteration, the algorithm stops unless $t = 0$. During the second procedure, each node $v$ is associated with the colors of the set of edges that points to it, $\langle n_t(e) | t(e) = v \rangle$, where $n_t$ in each set are listed in an increasing fashion and $t(e)$ is the target node of the edge $e$. Similarly to the ISCVs, we identify all the unique sets and assign to each one a new color, from which we update the coloring of the nodes. If after this update there are not new colors compared with the last iteration, the algorithm stops and the balanced coloring is obtained. 

Figure~\ref{fig:boldi_aldis}(b) shows the partitioning of a small graph containing five nodes according to Aldis'\cite{Aldis} algorithm. Initially, all nodes are assigned to the same class as well as the edges. Starting from iteration $t = 0$, the pairs $( n_0(e), n_0(s(e)) )$ are associated to each edge, which in this case is always $(1,1)$. In this step, all edges stay in the same class and coloring update does not introduce any new color. However, the algorithm does not stop because $t = 0$. Still at $t = 0$, the set $\langle n_0(e) | t(e) = v \rangle$ is associated to each node $v$. From this, there are two unique sets and we assign to each one of them a color (in this case, $\langle 1 \rangle \rightarrow \text{red}(1)$, $\langle 1, 1 \rangle \rightarrow \text{blue}(2)$) and we update the nodes color according to these labels. After that, we go to the next iterations $t = 1, 2, ...$ repeating the same procedures until no further refinement is possible, like displayed in Fig.~\ref{fig:boldi_aldis}(b), and the balanced coloring is reached.

Here we stress that this algorithm allows for both the set of nodes and the set of edges of the network to have an initial coloring different from the trivial (all nodes with the same color), since the two-procedure implementation guarantees that the algorithm does not depend on the initial conditions. 
Furthermore, as compared to the Kamei-Cock algorithm, the advantage of Aldis' approach is that it does not need to use a matrix to store the input set vectors, avoiding any problem regarding the sparse structure of the ISCVs. However, the algorithm exhibits an inferior runtime complexity than the KC coloring algorithm \cite{KameiCock} (See table~\ref{tab:perf}).

\section{Measuring the Relative Performance of the Algorithms}
\label{sec:comparison}

By now, we have presented four methods to identify cluster synchrony in information-processing networks. Within these four, we argue that the method proposed in this work, which we call fast fibration partitioning (FFP), is the most efficient algorithm so far, being capable to handle efficiently with both computational memory and time resources. 
As we have described in the last sections, both the FFP and KC algorithms are based on the refinement paradigm. However, due to the minimization nature of the problem, it is possible to design several refinement routines that require different time and space complexities. This is a consequence of the greatest fixed point theorem, demonstrated by Tarski in 1955 \cite{Tarski1955}. 
This theorem states that we can define the refinement operation by a monotonic function $g$ that splits every unbalanced (or unstable) class into several disjoint balanced classes. Being $g(\bar{P})$ a function of a partition $\bar{P}$ over $V$, the theorem states that when the greatest fixed point of the function $g$ is reached, meaning the minimum number of classes within $\bar{P}$ such that $g(\bar{P}) = \bar{P}$, then the resulting partition $\bar{P}$ represents the one containing the minimal number of classes over the given sets $V$ that it is stable with respect to the relation $E \subseteq V \times V$. As showed in \cite{PTarjan1985}, the general approach through a refinement paradigm can be defined as the following steps:
\begin{enumerate}
  \item Define the partition $\bar{P}$ through an initial partition $\bar{P_0}$ over $V$;
  \item Refine the current partition: $\bar{P} \leftarrow g(\bar{P})$;
  \item If the partition do not change, the algorithm stops. Otherwise, go to step 2.
\end{enumerate}

\begin{figure}[t]
  \centering
  \includegraphics[scale=0.36]{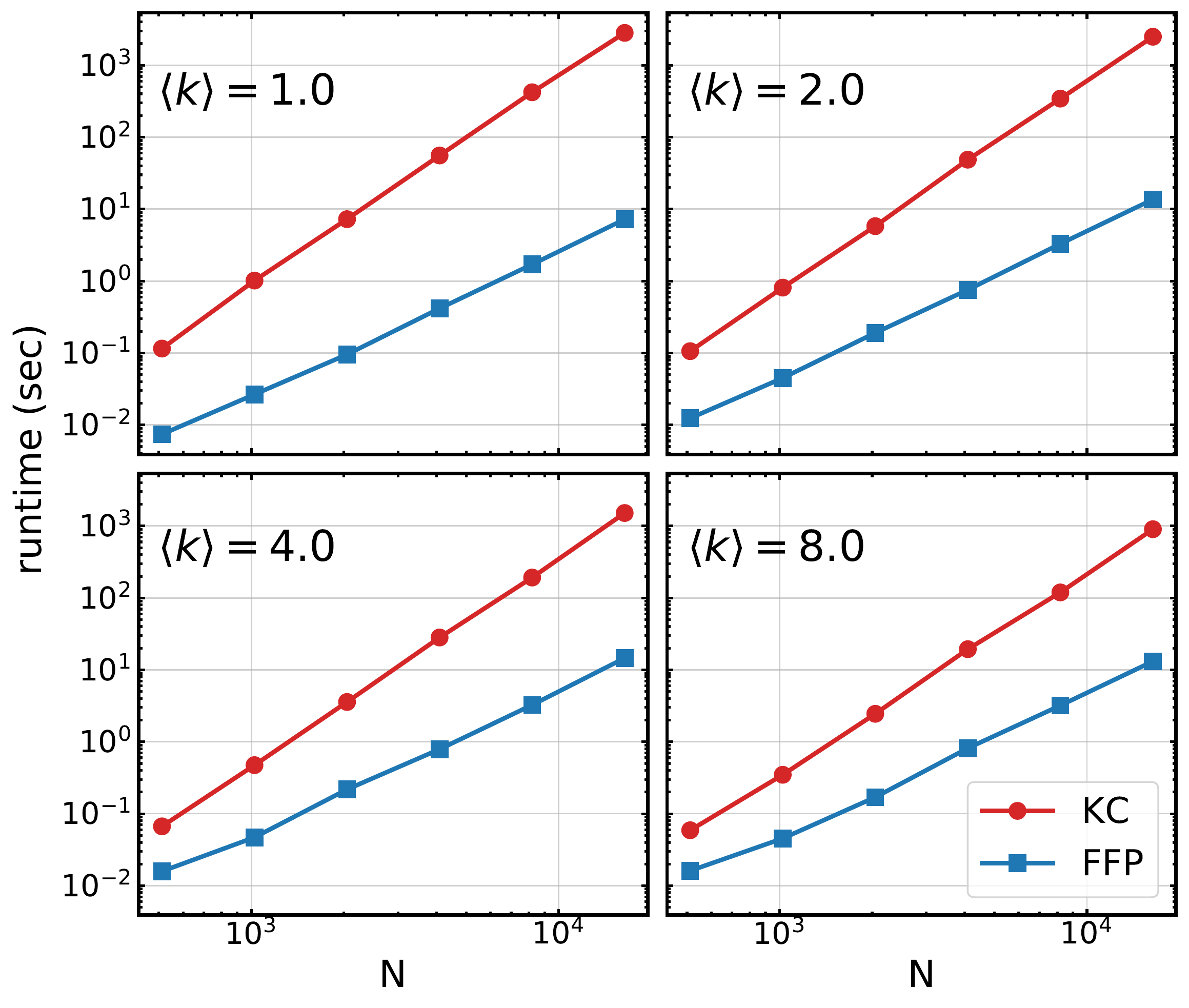}
  \caption{\textbf{Comparison between the runtime variation of the KC (red circles) and FFP (blue squares) algorithms as function of the size $N$ of Erdös-Renyi directed networks.} Here, networks have mean degrees $\langle k \rangle = 1$ (a), $2$ (b), $4$ (c), and $8$ (d). The symbols correspond to averages calculated over 20 network realizations and the size $N$ ranges from 512 to 16384 nodes. In all curves, the errors are smaller than the symbols. As $\langle k \rangle$ increases, {\it i. e.}, the network becomes denser (and the symmetries become trivial), the performance of the FFP algorithm is decreased, as expected from its time complexity $\mathcal{O}(M\log N)$, while the KC's performance improves, as it requires less iterations to find the trivial partitioning. However, this improvement is bounded as the network loses its symmetries, for which the large gap between both performances remains clear, where the FFP method is faster than the coloring algorithm.}
  \label{fig:singlelayer}
\end{figure}

Through this approach, it is possible to construct implementations either with runtime $\mathcal{O}(N^2\log N)$ or $\mathcal{O}(M\log N)$, where $N$ and $M$ represents, respectively, the number of nodes and the number of edges of the network. Considering graph fibrations, we can define the splitting functions both for the Kamei-Cock coloring (KC) and the fast fibration partitioning algorithm (FFP) as $s_{coloring}$ and $s_{input}$, respectively. 
The $s_{input}$ is the splitting function defined earlier in this work, while $s_{coloring}$ is a similar function also respecting the input set stability, but implemented according to the mechanism of the KC algorithm. As previously discussed in section~\ref{sec:fastfiber}, the FFP algorithm, with the $s_{input}$ splitting function, performs all refinement operations with runtime equal to $\mathcal{O}(M\log N)$. On the other hand, the KC algorithm, through its $s_{coloring}$ splitting function, have an overall runtime of at least $\mathcal{O}(N^2\log N)$, as we show as follows.

In the Kamei-Cock algorithm, for each step the algorithm verifies if each class is balanced by comparing the ISCVs of each node inside the current classes. In that manner, every node has its ISCVs checked and compared with the nodes sharing its same class. These comparisons and the splitting of unbalanced classes demand a runtime which grows with the size of the matrix of ISCVs, since each node has its ISCV verified. 
We notice that the matrix containing all ISCVs has $K^{i} \times K_{type} \times N$ entries, where $K^{i}$ corresponds to the number of colors at iteration $i$ and $K_{type}$ is equal to the number of edge types within the network. Since the refinement process allows a maximum number of $N$ colors for the worst-case scenario, the time complexity of the Kamei-Cock must be at least quadratic, which is the same order of the space resources necessary to its implementation.
Still considering the worst-case scenario where the final coloring contains $N$ colors, we can verify that to refine an initial coloring, where all nodes belong to the same class, into the final one with $N$ colors we need an order of $\log N$ iterations. For instance, if we start the refinement with a single class containing all nodes of the network, and for each iteration each class is splitted into two new classes with half the size of the splitted class, we expect exactly $\log N$ steps to obtain $N$ classes. 
Therefore, the time complexity of the Kamei-Cock algorithm is $\mathcal{O}(\alpha N^2\log N)$, where $\alpha = K_{type}$ representing the number of types of edges within the network.


To verify these considerations, we performed the comparisons between both algorithms, KC and FFP for two different cases in Erdös-Renyi networks. In general, fibration symmetries are rarely found in these networks, which allows us to assess the runtime of the methods in their worst-case scenarios. 
In Figs.~\ref{fig:singlelayer}(a)-(d) we show the runtime variation of the KC and FFP algorithms as function of the size of the Erdös-Renyi networks generated with mean degrees $\langle k \rangle = 1$,$2$,$4$, and $8$, respectively. As expected, the FFP algorithm performs much better in sparse networks, with slightly decreasing performance for denser networks. Furthermore, for all cases the runtime performance of FFP overcomes the one of the KC algorithm. 
Considering the second case, in Fig.~\ref{fig:multilayer} we perform the same analysis but varying the number of edge types for multiplex random networks. As previously mentioned, the presence of several types of edges affects drastically the performance of the KC algorithm, while the performance for the FFP almost does not change. 
Since for each new edge type introduced in the network the KC algorithm doubles the size of the ISCV matrix, its performance becomes restrictive even for moderate sizes of multiplex networks. As shown in Fig.~\ref{fig:multilayer}, the runtime difference between the performance of the KC algorithm for $K_{type} = 1$ and for $K_{type} = 8$ is almost 200 seconds for a network of size $N = 2048$ and mean degree $\langle k \rangle = 1.0$, while this difference is about 0.02 seconds for the runtime of the FFP algorithm.

Therefore, we observe that the method proposed here, based on the algorithm of Paige and Tarjan\cite{PTarjan1985}, outperforms the runtime performance of the Kamei-Cock algorithm used in \cite{fibration2019}. Furthermore, since the fast fibration partitioning algorithm requires only linear memory resources, the method also overcomes the KC algorithm in space resources, since the latter has at least $\mathcal{O}(N^2)$ complexity. In table~\ref{tab:perf} we list the time and space complexities of each algorithm mentioned in this work.
Regarding the algorithms of Boldi-Vigna and Aldis, we provide only the comparison between their complexities with the FFP algorithm. As we observe in table~\ref{tab:perf}, the time complexity of Aldis' algorithm does not allow us to compare its performance since it has runtime~\cite{Aldis} $\mathcal{O}(N(M+N)^3)$ and, hence, it is very constrained to small graphs. 
The Boldi-Vigna algorithm introduced in~\cite{Boldi2002} with runtime $\mathcal{O}(N^2d\log N)$ was designed for demonstration purposes and for pratical applications the authors of the algorithm~\cite{Boldi2002} used the method of Cardon and Crochemore~\cite{Cardon1982}, from which the Paige-Tarjan algorithm used in this work is an improvement~\cite{Tarjan1987}.



\begin{figure}[t]
  \centering
  \includegraphics[scale=0.435]{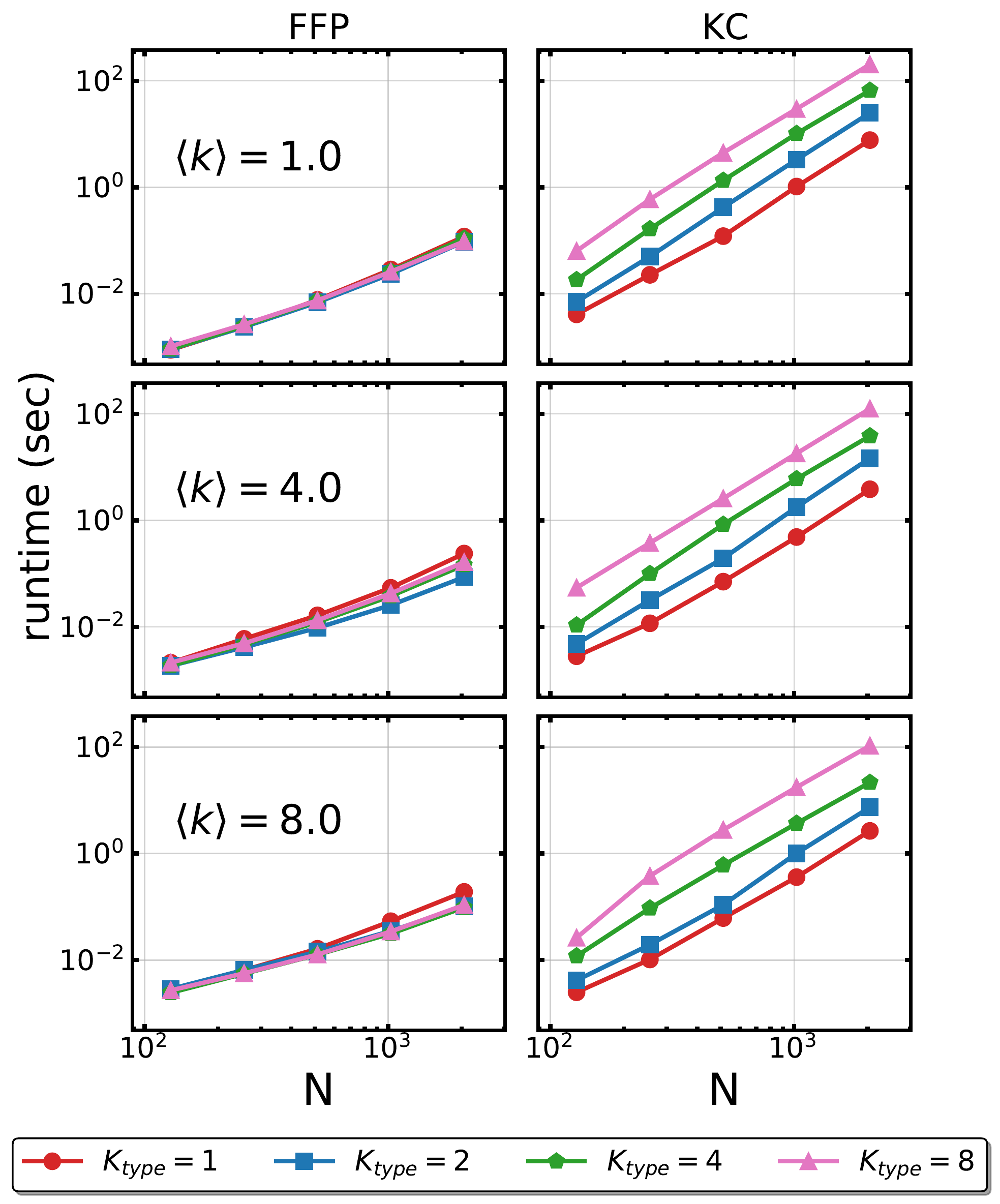}
  \caption{\textbf{Comparison between the runtime variation of the FFP (left column) and KC (right column) algorithms as a function of the size $N$ of Erdös-Renyi directed networks with different number of edge types.} Here, networks have $K_{types}=1$, $2$, $4$, and $8$ edge's types, and mean degrees $\langle k \rangle=1$, $4$, and $8$. The symbols correspond to averages calculated over 20 network realizations and the size $N$ ranges from 128 to 2048 nodes. In all curves, the errors are smaller than the symbols. Since the KC algorithm stores an ISCV matrix containing $K_{type} \times K \times N$ entries, where $K$ represents the number of colors, the variation of the number of edge types increases significantly the number of required operations. On the other hand, the FFP is much less sensitive to $K_{type}$ than the KC algorithm, being linearly dependent only with the size of the current pivot set $S$ during the iterations. Since the runtimes for the KC are very restrictive, we performed these comparisons only for very small networks.}
  \label{fig:multilayer}
\end{figure}



\setlength{\tabcolsep}{2pt}
\renewcommand{\arraystretch}{1.5}

\begin{table}
  \caption{\label{tab:perf} Summary of the time and space complexities of the algorithms mentioned in this work. In the algorithm from Boldi and Vigna, $d$ is the maximum in-degree in the network.}
  \begin{ruledtabular}
    \begin{tabular}{c|cc}
       & Time complexity & Space complexity \\
       \hline
      FFP & $\mathcal{O}  (M\log N)$ & $\mathcal{O}(M + N)$ \\
      KC \cite{KameiCock} &  $\mathcal{O}(N^2\log N)$ & $\mathcal{O}(N^2)$ \\
      Aldis \cite{Aldis} & $\mathcal{O}(N(M + N)^3)$ & $\mathcal{O}(M + N)$ \\
      Boldi and Vigna \cite{Boldi2002} & $\mathcal{O}(N^2d\log N)$ & $\mathcal{O}(Nd\log N)$ \\
    \end{tabular}
  \end{ruledtabular}

\end{table}

\section{conclusions}
\label{sec:conclusion}

The methods presented in this work represent the main collection of algorithms designed for the identification of fiber circuits. There is a remarkable presence of these circuits in several biological and non-biological networks, and, as showed in the very recent works \cite{MoroneNatComm2019,fibration2019,transistor2019}, each fiber leads to the synchronization between groups of nodes. 
Not only that, the symmetry breaking of these fibers can also lead to circuits analogous to logic computational units \cite{transistor2019}, affecting directly the functionality of the systems of which these circuits belong to. In summary, we have presented in this paper an optimal procedure to automatically identify fibrations in large networks. 
We showed that the balanced coloring method defined by the FFP algorithm outperforms, both in terms of runtime and space resources, other methods previously presented in the literature.

\section*{Code Availability}

Code for the Fast Fibration Partitioning algorithm is available at repository \url{https://github.com/makselab/} together with its documentation for proper use. We also provide the codes for the Kamei-Cock algorithm used in previous works with extended functionalities at \url{https://github.com/makselab/
fibrationSymmetries}.

\section*{Data Availability}

Data sharing is not applicable to this article as no new data were created or analyzed in this study, only synthetic data.

\section*{Acknowledgment}

This work was funded by NIH-NIBiB and NIH-NIMH through the Brain Initiative grant number R01 EB028157 and R01 EB022720. Financial support was also provided by the Brazilian agencies Cnpq, CAPES and FUNCAP, and also the National Institute of Science and Technology for Complex Systems.

\nocite{*}
\bibliography{sample}

\end{document}